# A Comparison of Maps Application Programming Interfaces

**Ana Isabel Fernandes, Miguel Goulão, Armanda Rodrigues**

CITI/FCT, Universidade Nova de Lisboa
Quinta da Torre, 2829 -516 CAPARICA, PORTUGAL
ai.fernandes@campus.fct.unl.pt, mgoul@fct.unl.pt, a.rodrigues@fct.unl.pt

**Abstract** The development of web applications that manipulate geo-referenced information is often supported by Application Programming Interfaces (APIs), allowing a fast development cycle for high quality applications. APIs can be used by programmers with different expertise levels and choosing an adequate API may have a dramatic impact on the productivity achieved by those programmers. Our goal is to compare maps APIs with respect to their usability. We compare three different APIs: the Google Maps JavaScript API, the ArcGIS API for JavaScript, and the OpenLayers JavaScript Mapping Library. Our comparison is supported by a set of software metrics and is performed in two orthogonal ways: the comparison of three implementations of the same system prototype, each using one of the APIs under scrutiny; the comparison of the APIs specifications. The main results of the study are related to the size of the APIs, with the Google API being significantly smaller than the others.

## 1 Introduction

The development of web applications that manipulate geo-referenced information is often supported by Application Programming Interfaces (APIs) that leverage specialized component frameworks for this domain. This form of reuse enables a rapid development cycle of high quality applications [1, 2].

An API is typically used by programmers, but it can also be used by domain experts who occasionally may play the role of programmers in the development process. When selecting among competing APIs, we search for an API that improves the productivity of programmers, through usability [3, 4]. Usability in APIs is related to facilitating the use of the set of actions which can be performed by the API [5]. This has driven several studies addressing the usability of APIs (e.g. [5-8]).

In this paper we compare three different maps APIs with respect to their usability. We chose three well-known APIs: Google Maps, OpenLayers and ArcGIS. These APIs represent a commercial, academic and GIS perspective of the maps APIs market, respectively. To the best of our knowledge, this is the first comparative study focusing on the usability of APIs for maps.

This paper is organized as follows. The current section introduces and motivates the problem of maps API comparison. Section 2 introduces the selected components for the test suite, the selected metrics for the usability evaluation and the developed tools for the results collection. Section 3 introduces the selected set of software metrics for the evaluation. Section 4 presents the resulting data obtained and an analysis of the results. In Section 5, we compare our work with the existing related work. Finally, in Section 6, conclusions are presented and future work is discussed.

## 2 Comparison setup

In this paper we use two different kinds of information sources: the APIs definitions, to assess their structural properties and their evolution over time, and three prototypes with similar functionality, built with the three chosen APIs.

### *2.1 Selected APIs and versions*

In this comparison we use the following versions of the APIs: Google 3.7 – 3.9, ArcGIS 2.0 – 3.1, and OpenLayers 2.3 – 2.12. The versions used were made available over the period of one year. The Google set of versions is not complete, as Google only keeps online details on the last three API versions.

### *2.2 Prototypes developed for facilitating the comparison*

We developed three prototypes in JavaScript supporting the same level of functionality, each with one of the selected APIs. The prototypes are used to assess how each of the APIs impacts the developed applications complexity. Figure 1 presents a screenshot of the Esri version of the prototype.

An important requirement for these prototypes was that they should be representative of typical functionalities to be found in maps applications. The identification of those functionalities combined the analysis of several well-known map applications and the syllabus of an introductory post-graduate course in Geographic Information Systems. The selected basic functionalities were: *zoom*, *full extent*, *pan*, *controllers*, *overview map*, *geo-referenced entities*, *information associated with entities* and *location search*. The functionality of each prototype was supported by each APIs implementation of a set of methods (presented in figure 7), representative of the case study application.

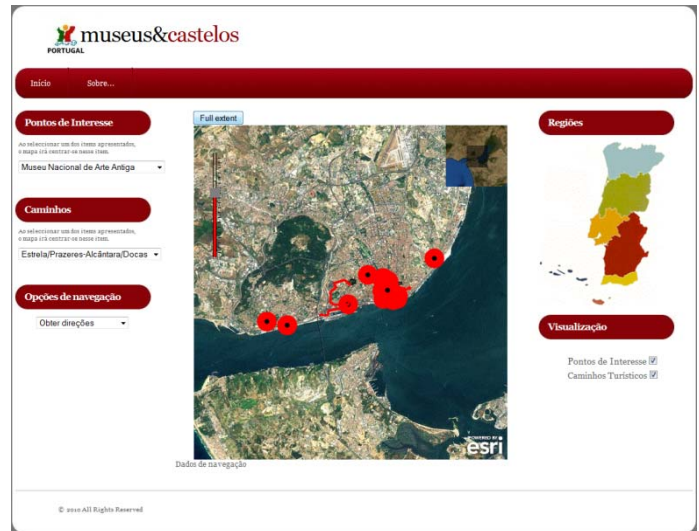

**Fig. 1.** Esri API based application

## 2.3 Metrics identification

We used the Goal-Question-Metric (GQM) approach [9] to support the quantitative comparison of the APIs. GQM starts with the identification of the research goals. Our *goals* are (G1) to characterize the evolution of each of the APIs, in order to reason about their stability, and (G2) to measure the actual use of the API in practice. To achieve each of these goals, we formulated research *questions* which, in order to be answered, required the identification of appropriate *metrics*. Table 1 summarizes the *goals*, *questions* and *metrics* used in this paper.

**Table 1.** GQM model

| Goal G1: Characterize the evolution of the APIs | |
|---|---|
| **Question:** | **Metric:** |
| **Q1.** What is the size of the API? | **NOMP N**umber of **O**bjects, **M**ethods and **P**roperties. |
| **Q2.** How many objects were deleted? | **NDO N**umber of **D**eleted **O**bjects. |
| **Q3.** How many objects were added? | **NAO N**umber of **A**dded **O**bjects. |
| **Q4.** How many objects were kept? | **NKO N**umber of **K**ept **O**bjects. |

| Goal G2: Assess the actual use of the API ||
| --- | --- |
| **Question:** | **Metric:** |
| **Q5.** How complex is the implementation of the functionality of the prototypes? | **APIUI API**'s **U**sage **I**ndex. |

### 2.3.1 API evolution characterization metrics

The complexity of an API may increase with the evolution of its functionality, in order to meet new user requirements [10, 11]. This can be assessed by using API size metrics and analyzing their variation from one version to the next. While the values for a single API version measure complexity, their variation over consecutive versions assess stability. API instability may imply extra costs for the evolution of applications. Table 2 describes the size metrics used in this study.

**Table 2.** API evolution characterization metrics

| Q1. What is the size of the API? ||
| --- | --- |
| **Metric** | **NOMP** – **N**umber of **O**bjects, **M**ethods and **P**roperties |
| **Definition** | Total number of objects, method and properties in the API. |
| **Comment** | Variations in this number can be used as a surrogate for variations in the API complexity. |
| **Q2. How many objects were deleted?** ||
| **Metric** | **NDO** – **N**umber of **D**eleted **O**bjects. |
| **Definition** | Total number of deleted objects. |
| **Comment** | Object deletion breaks the compatibility with API clients built for previous versions, contributing to increased evolution costs. |
| **Q3. How many objects were added?** ||
| **Metric** | **NAO** – **N**umber of **A**dded **O**bjects. |
| **Definition** | Total number of added objects. |
| **Comment** | New objects mean new functionalities offered by the API. |
| **Q4. How many objects were kept?** ||
| **Name** | **NKO** – **N**umber of **K**ept **O**bjects. |
| **Definition** | Total number of kept objects. |
| **Comment** | The "stable" part of the API. |

### 2.3.2 Metrics to assess the expressiveness of the API

The three prototypes implement the same functionality, so the API Usage Index metric, defined in table 3, can measure how many calls to constructors, functions and properties of the API are needed by each feature. In general, more expressive APIs will require fewer calls.

Table 3. Metrics for assessing the understandability of APIs

| Q5 - How complex is the implementation of the functionality of the prototypes? | |
|---|---|
| **Metric** | **APIUI** - **API U**sage **I**ndex |
| **Definition** | Number of calls to the API. $$C+F+P$$ $C$ - total calls to constructors; $F$ - total calls to functions; $P$ - total calls to properties. |
| **Comment** | A higher number of calls for achieving the same functionality may indicate less expressiveness from the API. |

## *2.4 Data collection*

Besides the development of the prototypes, we also implemented support for data collection from the prototypes and APIs.

Since the user interacts with the API by using its objects, data collection contemplated the information that the user has access to, such as objects, their methods and properties. To study the effects of the API usability, the implementation details of the APIs source code are irrelevant to the user, who observes them as black boxes.

## 3 Results and Discussion

### *3.1 Qualitative evaluation*

While developing the prototypes the most noticeable limitations were that:

- the Google API does not support the manipulation and management of layers
- the OpenLayers API does not support the geocoding process.

Concerning the remaining functionalities, all the APIs provided support for the features considered in this study and already mentioned in section 2.2.

## 3.2 Quantitative evaluation

### 3.2.1 Dimension of the APIs

The number of objects, methods and properties per object, are indicators of the dimension of the API. Figure 2 presents them for each of the compared APIs. The Google API has less than half of the objects of the OpenLayers API, while the ArcGIS API has an intermediate number of objects. Overall, the Google API is much smaller, in terms of objects, methods and properties, than the ArcGIS API which, in turn, is also much smaller than the OpenLayers API.

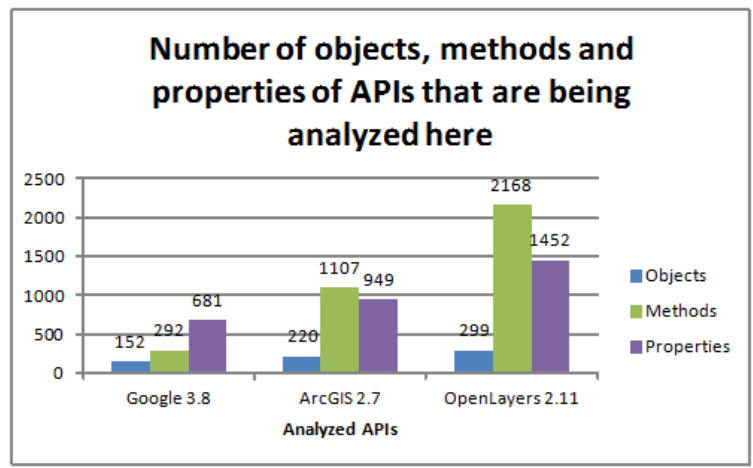

**Fig. 2.** Number of objects, methods and properties of APIs that are being analyzed here

### 3.2.2 Evolution of the APIs

Figures 3 through 8 present the evolution of the number of objects, methods and properties for each of the APIs, as well as the number of added, deleted and kept objects. The three APIs are growing, considering each of these elements.

The Google API is the only one with no decreases in the analyzed period (figure 3). No object deletions occur in these versions (figure 4).

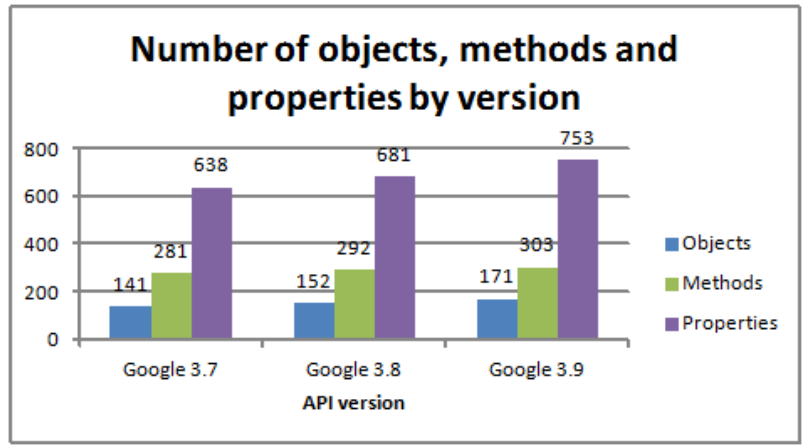

**Fig. 3.** Number of objects, methods and properties along the versions – Google

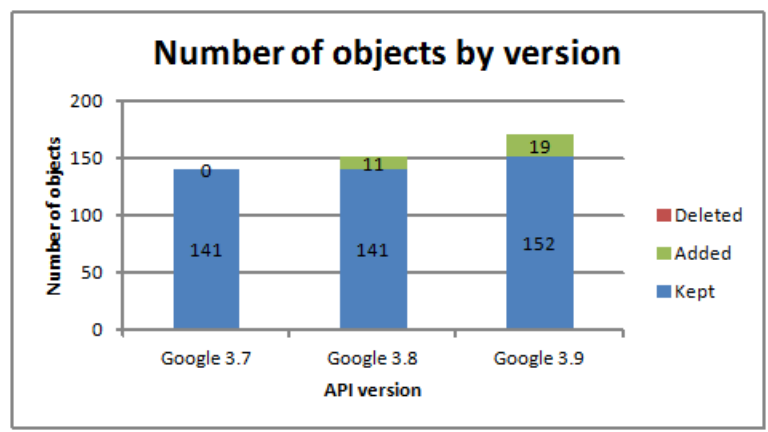

**Fig. 4.** Number of objects along the versions – Google

The Esri API suffered a decrease from versions 2.3 to 2.4 for objects, methods and properties. The number of methods in the API also decreases in version 3.0 (figure 5). Three of its versions have deleted objects (figure 6).

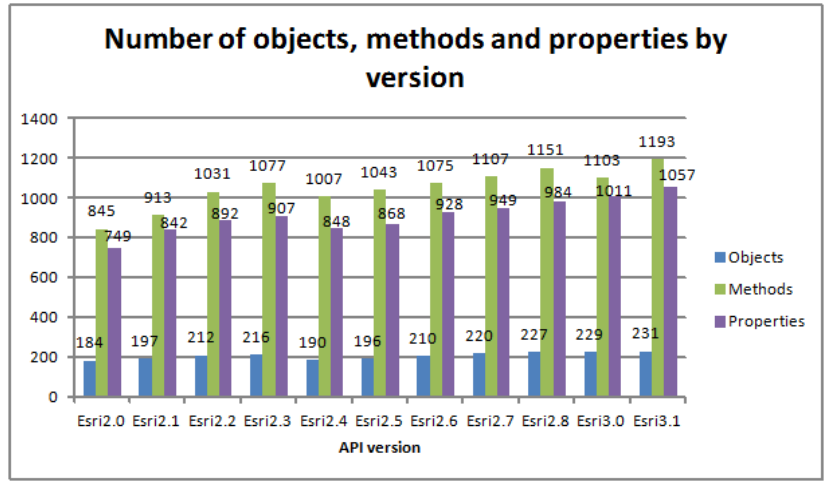

**Fig. 5.** Number of objects, methods and properties along the versions – ArcGIS

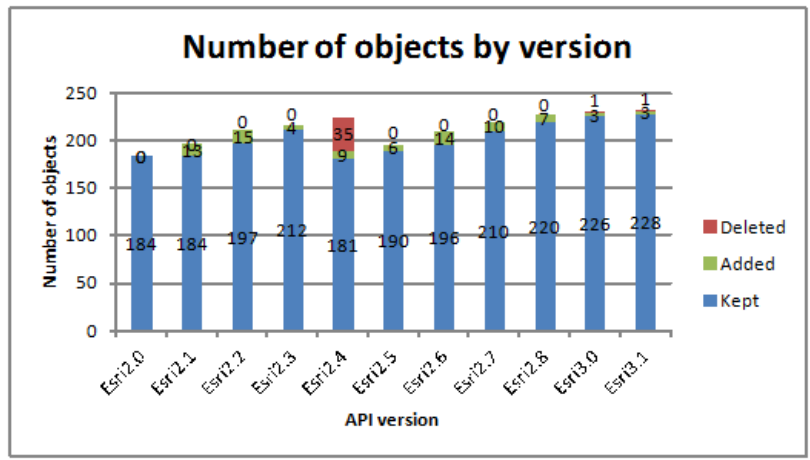

**Fig. 6.** Number of objects along the versions – ArcGIS

The OpenLayers API, suffered a decrease in objects, methods and properties in version 2.12 (figure 7). In four of its versions, there are deleted objects (figure 8).

The elimination of objects, methods and properties may impact the APIs users' productivity, in the evolution of their own maps applications, as they will have to learn how to cope with the new alternative support for the deleted elements and change their applications accordingly.

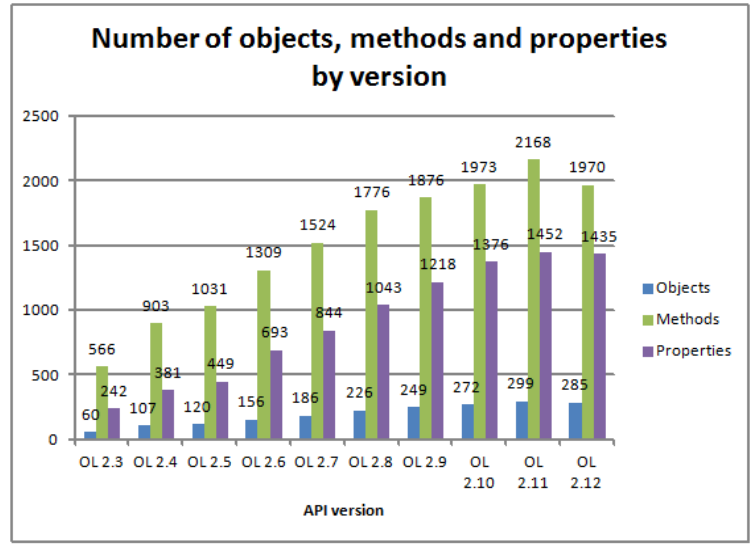

**Fig. 7.** Number of objects, methods and properties along the versions –OpenLayers

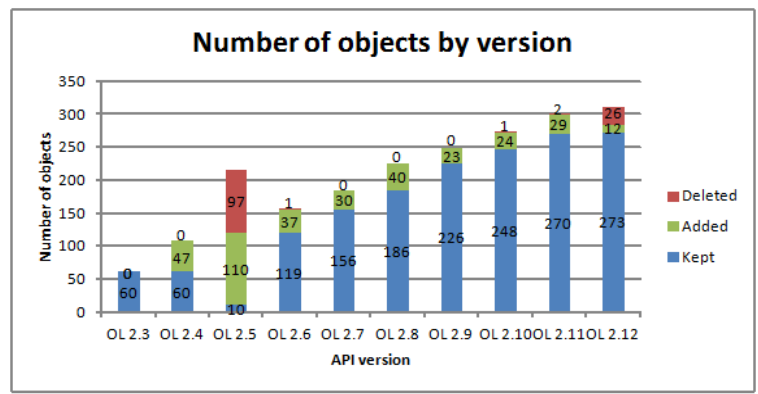

**Fig. 8.** Number of objects along the versions - OpenLayers

### 3.2.3 Evaluation of the actual API use

While in the previous section the metrics were collected from the APIs definitions, in this section we discuss a metric aimed at assessing the actual usage of an API – the API Usage Index (APIUI) – and collect it from our set of functionally equivalent prototypes.

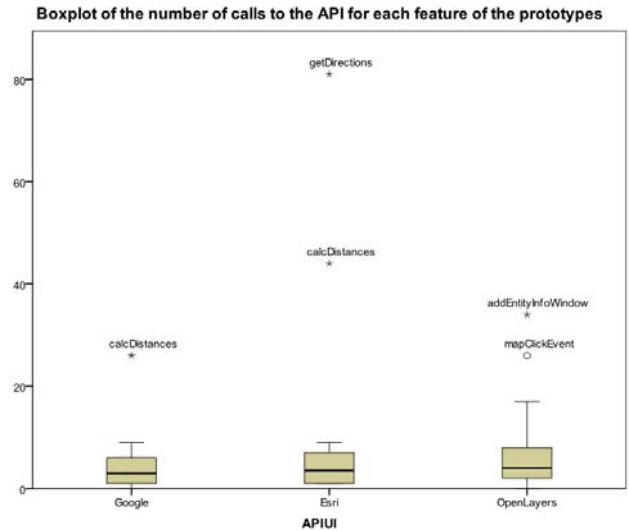

**Fig. 9.** Number of calls to the API

Figure 9 presents a boxplot with the distribution of APIUI, comparing how the 3 APIs were used in practice. The mean number of API calls to each operation (3 to Google API, 3.5 to Esri API and 4 to OpenLayers API) suggests that the methods of the developed prototype supported by the OpenLayers API need, in average, to make more calls to the API.

The outlier and extreme values in the figure indicate the methods that involved more calls to the API, in their respective prototypes. The *calcDistances* is an extreme value both for Google and the Esri API, while the *getDirections* method is also an extreme value for the Esri API. Furthermore, the OpenLayers prototype has an outlier value in the *mapClickEvent* method and an extreme value in the *addEntityInfoWindow* method.

In Figure 10, we present a different view on the same data: the number of necessary calls to each API, by prototype functionality. As expected, the Google prototype needs fewer calls to the API, while the OpenLayers prototype generally needs more calls per functionality.

The data is not normally distributed, as tested by a Shapiro-Wilk test (p-value < 0.001). In this case we use the Shapiro-Wilk test due to the sample size, which is smaller than 50. A Kruskal-Wallis test shows there is no statistically significant difference among the APIs (p-value >= 0.05). As such, we cannot assume there is a significant difference in the number of API calls when using the different APIs.

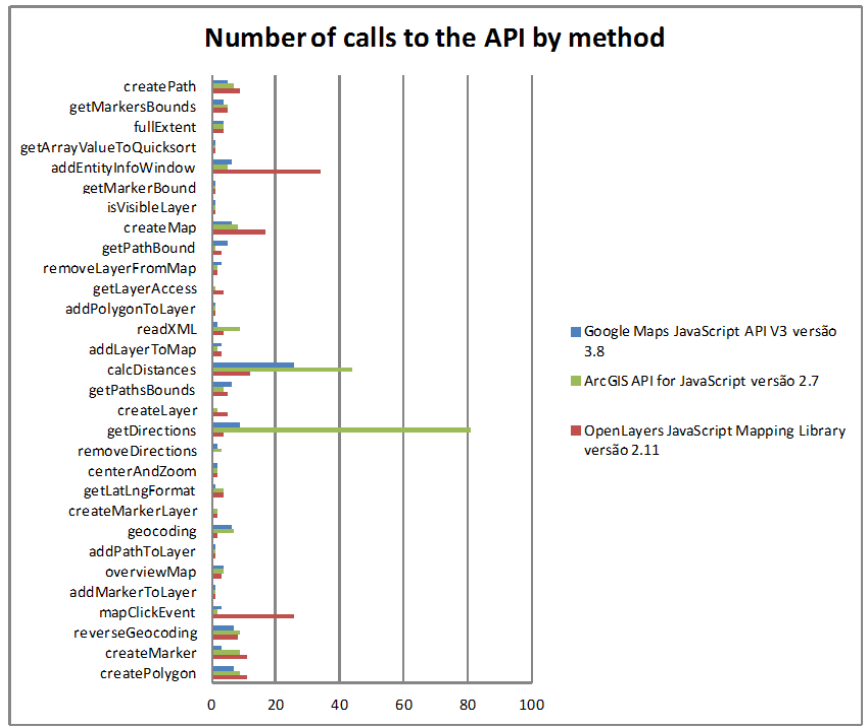

**Fig. 10.** Number of calls to the API

## 3.2 Discussion

The Google API is significantly smaller than the other two APIs. In turn, the OpenLayers API is the largest. A larger API may create more difficulties in terms of learning, understanding and usage, since the complexity of an API is frequently correlated to its dimension.

From the data obtained from the usage evaluation metrics, we could conclude that the prototype developed with the OpenLayers API (the largest), involved more calls to the API, while the prototype developed with the Google API (the smallest) involved less API calls. The OpenLayers API, because of its dimension, needs a larger effort in the choice of the most adequate objects for each task.

In a more reduced API, like the Google API, the provided objects may allow a higher level of abstraction, which facilitates the learning and usage of an API. The Google API based prototype has a reduced number of calls to the API, in comparison with the OpenLayers API based prototype, although the difference is not statistically significant.

Finally, from the data obtained from the analysis of the several versions of each API, and considering the laws of software evolution [10, 11], we confirmed that the APIs are evolving, and that the understandability of consecutive versions is stable. However, in terms of retro-compatibility, the Google API is the only one with no removals of objects between the analyzed versions. Migrating applications from one version to the next is facilitated in this API, when compared to what happens with the OpenLayers and Esri APIs.

## 4 Related Work

Regarding the domain of the APIs, this work differs from existing work in the area of usability of APIs, since they do not consider the category of maps APIs. The methodology used for our usability comparison also differs from what we commonly find in most API usability studies (e.g. [5-8]). The most common method for evaluating the usability of APIs involves usability testing in a laboratory environment, where a set of API users are asked to perform a set of tasks using a particular API. These tests heavily rely on the availability of API users. In this study, we conducted an evaluation through data extraction from the source code of prototypes and from the specification of the studied APIs.

## 5 Conclusions and future work

In this paper we compared three different maps APIs, from a usability perspective. To support this comparison, we used two different data sets: the APIs definition and three prototypes implementing similar functionality on the chosen APIs.

The qualitative analysis was based on the construction of the prototypes. The implementation allowed the identification of the following limitations: the Google API does not allow layer management and the OpenLayers API does not support geocoding.

Concerning the quantitative analysis, we selected appropriate metrics to compare the usability of the different APIs. From the collected data, we found that the Google API is considerably smaller than the other two APIs, while the OpenLayers API is the largest. In principle, a larger API should imply a greater difficulty in learning, comprehension and use, since complexity is often correlated with the API's dimension.

Finally, from the data obtained from the analysis of the several versions of each API, and considering the laws of software evolution [10, 11], we conclude that their understandability remains fairly stable in all analyzed aspects but API size. A challenge was identified concerning stability: the removal of objects between versions, in the OpenLayers API and the Esri API. This may break the compatibility of applications using those APIs, when migrating to more recent versions of the APIs.

As future work, we expect to explore repositories with map applications so that we can get a more reliable picture of the actual use of these APIs, in a professional context. It would also be interesting to combine process data, such as effort (measured as the time a user takes to perform a task), and the rate of defects introduced by using the APIs. Finally, there is also the possibility of conducting a study to evaluate the usability of Maps APIs in mobile devices.

## References

1. Charles W. Krueger. Software reuse. ACM Comput. Surv., 24(2):131–183, June 1992.

2. W. B Frakes e Kyo Kang. Software reuse research: status and future. IEEE Transactions on Software Engineering, 31(7):529– 536, July 2005

3. John M. Daughtry, Umer Farooq, Brad A. Myers, e Jeffrey Stylos. API usability: report on special interest group at CHI. SIGSOFT Softw. Eng. Notes, 34(4):27–29, July 2009.

4. Cleidson R. B. de Souza, David Redmiles, Li-Te Cheng, David Millen, e John Patterson. Sometimes you need to see through walls: a field study of application programming interfaces. In Proceedings of the 2004 ACM conference on Computer supported cooperative work, CSCW '04, pp. 63–71, New York, NY, USA, 2004. ACM.

5. Steven Clarke. Measuring API usability. Dr. Dobb's Journal, 29:S6–S9, 2004.

6. Martin P. Robillard. What makes APIs hard to learn? answers from developers. IEEE Softw., 26(6):27–34, November 2009.

7. Jeffrey Stylos e Steven Clarke. Usability implications of requiring parameters in objects' constructors. In Proceedings of the 29th international conference on Software Engineering, ICSE '07, pp. 529–539, Washington, DC, USA, 2007. IEEE Computer Society.

8. Umer Farooq, Leon Welicki, e Dieter Zirkler. API usability peer reviews: a method for evaluating the usability of application programming interfaces. In Proceedings of the 28th international conference on Human factors in computing systems, CHI '10, pp. 2327–2336, New York, NY, USA, 2010. ACM.

9. Victor R. Basili, Gianluigi Caldiera & Dieter H. Rombach. Goal Question Metric Paradigm. In John J. Marciniak, editor, Encyclo-pedia of Software Engineering, volume 1, pages 469–476. John Wiley & Sons, 1994.

10. Meir M. Lehman. Programs, cities, students, limits to growth? Programming Methodology, pp. 42–62, 1978. Inaugural Lecture.

11. Meir .M. Lehman. On understanding laws, evolution, and conservation in the large-program life cycle. Journal of Systems and Software, 1(0):213–221, 1979.